\documentclass[twocolumn,aps,showpacs,preprintnumbers,amsmath,amssymb]{revtex4}
\usepackage{epsf}	

\begin{document}
\title{Parametric Driving of Dark Solitons in Atomic Bose-Einstein Condensates}
\author{N.P. Proukakis$^\dag$, N.G. Parker$^\dag$, C.F. Barenghi$^\ddag$, and C.S. Adams $^\dag$}
\affiliation{$^\dag$ Department of Physics, University of Durham, South Road, Durham, DH1
3LE, United Kingdom}
\affiliation{$^\ddag$ School of Mathematics and Statistics, University of Newcastle, Newcastle upon
Tyne, NE1 7RU, United Kingdom}

\begin{abstract}

A dark soliton oscillating in an elongated harmonically-confined atomic Bose-Einstein condensate
 continuously exchanges energy with the  
sound field.  Periodic optical `paddles' are employed to controllably enhance 
the sound density and transfer energy 
to the soliton, analogous to parametric driving.  
In the absence of damping, the amplitude of the soliton oscillations can be
dramatically reduced,
whereas with damping,
a driven soliton equilibrates as a {\it stable} soliton with lower energy, 
thereby extending the soliton lifetime up to the lifetime of the condensate.

\end{abstract}
\maketitle

Dark solitons \cite{Kivshar} are an important manifestation of the intrinsic nonlinearity of a 
system and arise in diverse systems such as optical fibers \cite{ds_1},
waveguides \cite{ds_2}, surfaces of shallow liquids \cite{water}, magnetic films \cite{ds_5}, 
and atomic Bose-Einstein Condensates (BECs) \cite{BEC_sol}.
Dark solitons are known to be dynamically unstable in higher than 
one-dimensional (1D) manifolds (e.g. snake instability in 3D systems leading to a 
decay into vortex rings \cite{snake_0,dynamical_inst}).  Solitons in 
1D geometries
experience other instabilities, whose nature depends on the details of the system:
For example, dark solitons in optical media are prone to nonlinearity-induced changes 
in the refractive index \cite{Kivshar}, whereas in harmonically trapped
atomic BECs they experience dynamical instabilities due to the longitudinal
confinement \cite{Busch,Muryshev2,soliton_osc1,soliton_osc2,parker1,parker2,brazhnyi},
as well as thermodynamic \cite{thermo_inst} and quantum \cite{quantum_fluct} effects.
Instabilities lead to dissipation, which manifests itself in the emission of radiation.
Compensation against dissipative losses by parametric driving
has been demonstrated in some of the above media
\cite{water,water_theory}.
The aim of this Letter is to discuss this
effect in the context of atomic BECs.

In atomic gases, the snake instability \cite{dynamical_inst} can be suppressed in elongated,
quasi-one-dimensional (quasi-1D) geometries
\cite{quasi_1D}, and thermal instabilities 
 are minimized at very low temperatures $T \ll T_{c}$
(where $T_{c}$ is the BEC transition temperature).
In this limit, a dark soliton oscillating in a harmonically confined BEC 
continuously emits radiation (in the form of sound waves) 
due to the inhomogeneous background density 
\cite{Busch}.  The sound remains confined and re-interacts with the 
soliton, leading to periodic oscillations of the soliton energy \cite{parker1}.
In this Letter we propose the 
controlled amplification of the background sound field, and illustrate the 
resulting transfer of energy into the soliton, in close analogy to 
established parametric driving techniques \cite{water}.
Energy is pumped into the sound field via
periodically-modulated `paddles', located towards the 
condensate edge (Fig. 1). If the drive frequency is nearly resonant
with the soliton oscillation  frequency, one observes significant energy transfer to the 
soliton. 
In the absence of dissipation, this leads to a dramatic reduction in
the amplitude of the soliton oscillations.
Under dissipative conditions, the damped
soliton equilibrates as a {\it stable} soliton with lower energy, with its lifetime extended up to
 the condensate lifetime. 
Moreover, suitable
engineering of the phase of the driving field (relative to the soliton oscillations)
can maintain the soliton energy at its initial value
 for times singificantly longer than the undriven soliton lifetime.

\begin{figure}[b]
\epsfxsize=8cm \centerline{\epsffile{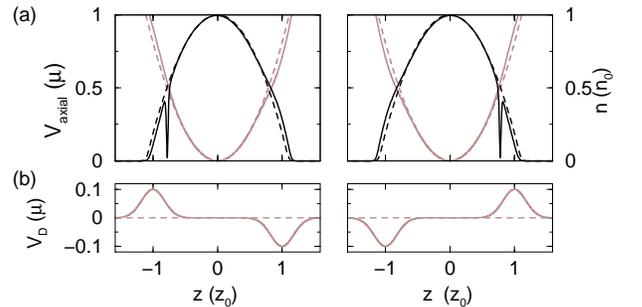}}
\caption{Schematic of parametric driving: (a) Total axial potential (grey lines, left axis) and
density (black lines, right axis) of perturbed harmonic trap with propagating dark soliton (solid lines),
at two times (left/right plots) corresponding to maximum drive
amplitudes. (b) Corresponding drive potentials (black, $\alpha >0$). Dashed grey lines denote density of
unperturbed harmonic trap, while dashed black lines indicate corresponding density.
}
\end{figure}


Our analysis is based on the cylindrically-symmetric 3D Gross-Pitaevskii Equation 
(GPE) describing the evolution of
the macroscopic order parameter  $\psi(\rho,z)$ of an elongated 3D atomic BEC 
\begin{eqnarray}
i\hbar \frac{\partial \psi}{\partial
t}=-\frac{\hbar^2}{2m}\nabla^2 \psi+V\psi+g|\psi|^2\psi-\mu\psi.
\end{eqnarray}
where $m$ is the atomic mass, $V=V_{\rm T}({\bf r})+V_{\rm D}({\bf r})$, $V_{\rm T}({\bf r})=(m/2)(\omega_{z}^{2} z^{2} + 
\omega_{\perp}^{2}\rho^{2})$ is the harmonic confining
potential of longitudinal (transverse) frequency $\omega_{z}$
($\omega_{\perp}$), where $\omega_{z} \ll \omega_{\perp}$, and $V_{\rm D}({\bf r})$ is the drive
potential (Eq. (3)). The
nonlinearity arises from atomic interactions yielding a scattering
amplitude $g=4\pi\hbar^2a/m$, where $a$ is the {\it s}-wave
scattering length. In this work, $a>0$, i.e. effective repulsive
atomic interactions. The chemical potential  is given by $\mu=gn_{0}$,
where $n_{0}$ is the peak atomic density.

Dark soliton solutions are supported by the 1D form of Eq.
(1) in the absence of external
confinement ($V=0$). 
On a uniform
background density $n$, a dark soliton with speed $v$ and position
$(z-vt)$ has the form,
\begin{eqnarray}
\psi(z,t)=\sqrt{n}e^{-i(\mu/\hbar)t} \left\{ \beta \tanh \left[
\beta
\frac{\left(z-vt\right)}{\xi}\right]+i\left(\frac{v}{c}\right) \right\}
\end{eqnarray}
where  $\beta=\sqrt{1-\left(v/c\right)^2}$, and the healing length
$\xi=\hbar/\sqrt{\mu m}$ characterises the soliton width.
The soliton speed
$v/c=\sqrt{1-(n_d/n)}=\cos(S/2)$ depends on  the total phase slip $S$
across the centre and the soliton depth $n_d$ (with
respect to the background density), with the limiting value set by
 the Bogoliubov speed of sound
$c=\sqrt{\mu/m}$.  The
energy of the unperturbed dark soliton of Eq. (2) is given by
$E_{\rm s}^{0}=(4/3)\hbar c n (1-(v/c)^{2})^{3/2}$. The drive
potential
\begin{equation}
V_{\rm D}=\alpha \sin(\omega_{\rm D}t) \left[e^{-(z+z_0)^2/w_{0}^{2}}-e^{-(z-z_0)^2/w_{0}^{2}} \right]
\end{equation} 
consists of two periodically modulated gaussian `paddles', 
with amplitude $\alpha$, at positions
 $\pm z_0$, oscillating in anti-phase at a {\it fixed} frequency $\omega_{\rm D}$,
close to the soliton frequency (Fig. 1).
Such a set-up could be created by time-dependent red and blue detuned
laser beams with beam waist $w_{0}$.

In our work, the dark soliton
is defined as the density deviation from the unperturbed density (in the absence of the soliton) 
within the `soliton region'  $R$: $\left[ z_{\rm s} - 5\xi, z_{\rm s} + 5 \xi\right]$, 
where $z_{\rm s}$ the instantaneous position of the local density minimum.
Such a `perturbed' soliton includes sound excitations located within the soliton
region at any time.
The motion and stability of a quasi-1D dark soliton is well parametrized by its 
energy. In order to facilitate a
direct comparison of the quasi-1D soliton decay to the analytical homogeneous 1D soliton energy,
the soliton dynamics are parametrized in terms of the
`on-axis' ($\rho=0$) soliton energy $E_{\rm s}=\int_{R} 
\left\{ \varepsilon[\psi(0,z)]- \varepsilon[\psi_{\rm bg}(0,z)] \right\} dz$,
where  $\varepsilon(\psi)=\hbar^2/(2m)\left|\nabla\psi\right|^2
+V\left|\psi\right|^2+(g/2)\left|\psi\right|^4$ and $\varepsilon[\psi_{\rm bg}(0,z)]$ is the
corresponding energy contribution of the background fluid \cite{parker1,parker2,carr}.

\begin{figure}[t]
\epsfxsize=8cm \centerline{\epsffile{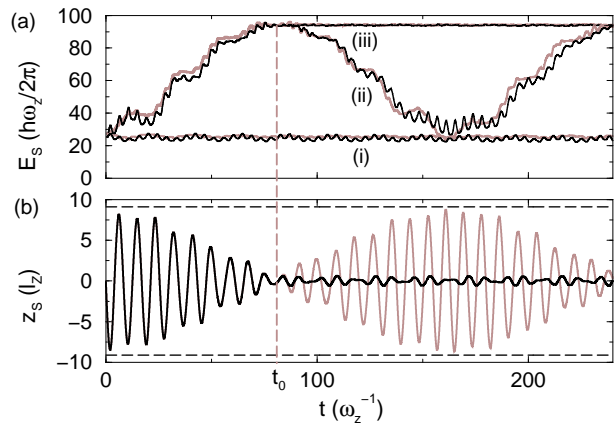}}
\caption{`On axis' quasi-1D soliton energy for (i) undriven case, 
(ii) continuous driving and (iii)
driving switched off at $t_{0}=80\omega_{z}^{-1}$, based on
simulations of the 3D cylindrically-symmetric GPE (black lines) and 1D GPE (grey lines)
for a soliton with initial speed $v=0.75c$. 
(b) Longitudinal soliton oscillations 
with continuous driving (grey
line), and driving switched off at $t_{0}$ (vertical grey line) under the 3D GPE. Dashed lines
indicate corresponding amplitude in absence of driving. The trap
strength is determined from the chemical potential of the system:
Quasi-1D: $\mu_{\rm 3D}= 8 \hbar \bar{\omega}$ where $\bar{\omega}=(\omega_{z}\omega_{\perp}^{2})^{1/3}$
and $\omega_{\perp}/\omega_{z}=250$.
Pure 1D: $\mu_{\rm 1D} = 70 \hbar \omega_{z}$ for which the 1D density matches the quasi-1D
 longitudinal density.
Drive parameters: $\omega_{\rm D}=0.98 \omega_{\rm sol}$, $\alpha=0.1 \mu_{1D}$, $w_{0}=3.2 l_{z}$
and $z_0=10.7 l_{z}$ where $l_{z}=\sqrt{\hbar/(m \omega_{z})}$ the longitudinal harmonic oscillator
length.}
\end{figure}

{\bf Dissipationless Regime:}
Consider first the case of
no dissipation (Eq. (1)). 
In the absence of $V_{\rm D}$, the soliton oscillates at
$\omega_{\rm sol}=\omega_{z}/\sqrt{2}$ 
\cite{Busch,Muryshev2,soliton_osc1,soliton_osc2,parker1,parker2,brazhnyi,thermo_inst}, 
emitting sound waves which oscillate at the
trap frequency $\omega_{z}$. This frequency mismatch means that the soliton
propagates through a periodically modulated background
density, 
leading to a {\em weak} periodic modulation of the soliton energy (Fig. 2(a), curves (i)).
The amplitude of this modulation is
enhanced by the coupling between longitudinal and
transverse degrees of freedom \cite{soliton_osc1,parker1}. 


To demonstrate substantial
energy transfer into the soliton, we start with a low energy
shallow soliton (speed $v_{0}=0.75c$ at $z=0$). Applying the drive potential induces an {\em additional} periodic background
density modulation and a time-dependence in the soliton
oscillation frequency, which is found to vary by no more than $10\%$ around its unperturbed value $\omega_{\rm sol}$. 
As a result, the relative phase between
the drive and the soliton oscillations, which determines the direction of energy flow between soliton and sound,
 becomes time-dependent.
Beginning with the drive out-of-phase with the soliton oscillations, the soliton initially acquires energy,
up to time $t_{0}$, after which it begins to lose energy,
and the cycle repeats, as shown by curves (ii) in Fig. 2(a).
This figure illustrates the good agreement between the 3D `on-axis' energy (black lines)
and the corresponding energy of the pure 1D simulations (grey lines) (the 3D results feature
an additional small amplitude oscillation
due to longitudinal-transverse coupling). The  corresponding beating in the soliton 
oscillation amplitude is shown in Fig.~2(b) (grey line). 
This beating effect can be visualized as
the periodic cycling between the initial low energy dark soliton, and the 
nearly
stationary high energy soliton. This
 picture is analogous to the
cycling of a driven condensate between the `no-vortex' and
`single-vortex' configurations \cite{keith}.

A soliton of higher energy can be created by
removing the drive potential after a certain pumping time.
In general, stopping the drive at an arbitrary point in the
energy gain-loss cycle, will lead to stabilization {\it around} that energy,
with associated residual oscillations. 
In order to create a nearly stationary soliton (black line in Fig. 2(b)), one
must stop the drive when the soliton has acquired its maximum energy,
for which the energy oscillations are suppressed (curve (iii) in Fig. 2(a)).


The soliton dynamics depend rather sensitively on the parameters of the driving field
which
should be carefully optimized for these effects to be clearly
observable. Firstly, the pumping should take place outside of the
range of the soliton oscillations
(i.e. $z_0 > 9 l_{z}$ for $v_{0}=0.75c$). If this is not the case,
the soliton traverses the gaussian bumps, leading to
`dephasing' of emitted sound waves and subsequent decay of the
soliton \cite{parker2}. The transfer of energy between
the soliton and the sound field depends on the 
phase of the drive relative to the soliton oscillations, and hence on the
drive potential
 seen by the soliton at the extrema of its oscillatory
motion. This parameter depends in turn not only on the drive frequency $\omega_{\rm D}$, 
but also on the amplitude
of the potential modulation $\alpha$, the range of the potential
$w_{0}$, its location $z_{0}$, and the initial soliton speed $v_{0}$.  If
all but one of the above parameters are kept constant, then there
is a resonance around an optimum value of that parameter. This
resonance is illustrated for the drive frequency $\omega_{\rm D}$ by the
open circles in Fig.~3. Note that the maximum pumping does not
arise at $\omega_{\rm sol}$, due to the additional frequency modification
induced by the perturbing potential; however, the optimum frequency
is consistently found  to lie close to  the unperturbed
frequency.
Importantly, the width of the
resonance for which the transferred soliton energy reaches half
its maximum value (FWHM), is reasonably broad, of the order of
$10\%$ the soliton frequency. 
We have also investigated alternative schemes for pumping energy
into the system. For example, using one off-centre paddle, or inverting the sign in Eq.~(3) 
(i.e. $\alpha <0$), leads to a delayed and less efficient energy transfer, 
while periodically displacing the trap
 leads to no net increase in the soliton 
energy. 

\begin{figure}[t]
\epsfxsize=6cm \centerline{\epsffile{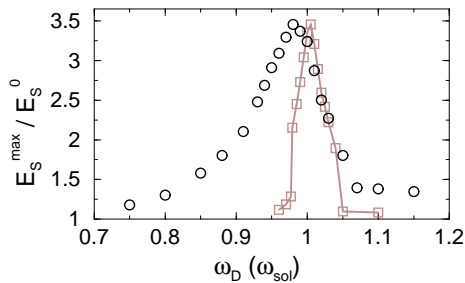}}
\caption{Ratio of maximum pumped energy $E_{\rm s}^{\rm max}$ to initial soliton energy 
$E_{\rm s}^{0}$ for a soliton with initial speed $v_{0}=0.75c$ 
in the presence of optimized driving as a function of drive frequency
(in units of the unperturbed soliton frequency $\omega_{\rm sol}=\omega_{\rm z}/\sqrt{2}$)
for (i) no damping (black circles),
or (ii) with damping $\gamma=5 \times 10^{-4}$ (grey squares). This value of $\gamma$ leads to the same 
soliton lifetime as for the undriven $0.3c$ soliton in 
Fig.~4.
Results based on pure 1D GPE for the case of Fig. 2.
}
\end{figure}

{\bf Dissipative Regime:}
In a realistic quasi-1D system 
featuring suppressed snake-instability \cite{dynamical_inst},
both the condensate and the 
soliton will be
prone to damping, e.g. due to the presence of a small thermal
cloud \cite{thermo_inst}. 
A first estimate into the effect of dissipation can
be obtained by introducting a phenomenological damping term
$\hbar \gamma \partial \psi / \partial t$ on the left hand side of Eq. (1). In the absence of
driving, this term leads to an approximately exponential decay of the
soliton energy, modified by the oscillatory motion of the soliton
(dashed lines in Fig. 4(a)).  Dissipation also damps both the sound field,
leading to a narrowing of the resonance in the drive frequency, and a shift of the resonant frequency towards the
unperturbed soliton frequency (grey data in Fig. 3). 

\begin{figure}[b]
\epsfxsize=8.5cm \centerline{\epsffile{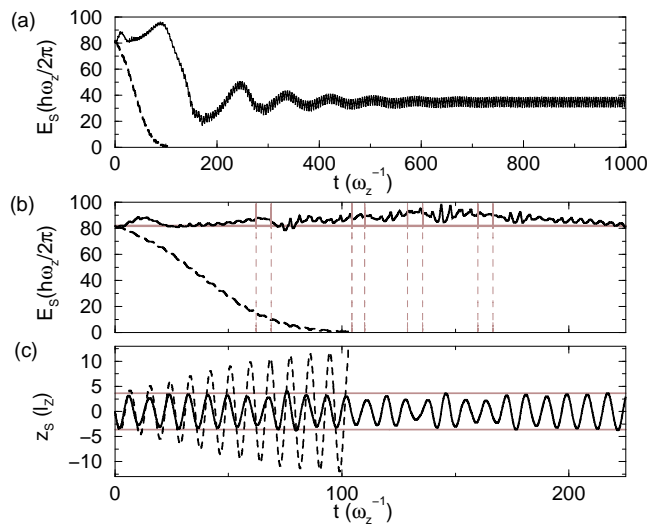}}
\caption{
(a) 
Soliton energy (initial speed $v_{0}=0.3c$) for a dissipative system ($\gamma =10^{-3}$) in the
presence (solid line) or absence (dashed) of continuous driving   
($\omega_{\rm D}=\omega_{\rm sol}$, $z_{0}=7.1l_{z}$). The driven soliton stabilizes at
$E \approx 35 \hbar \omega_{z}$, corresponding to $v_{0} \approx 0.7 c$.
(b) 
Energy of a $v_{0}=0.3c$ soliton with parametric driving (solid black), 
with a sequence of rephasing cycles 
during the periods between adjacent dashed grey lines.
Corresponding curves in the absence of driving are also shown 
for a dissipative($\gamma =10^{-3}$, dashed black) and a non-dissipative
$v_{0}=0.3c$ soliton (solid grey).
(c)
Driven (solid) and undriven (dashed) soliton trajectories in the presence of dissipation for case (b).
Horizontal grey lines indicate the corresponding undamped undriven {\it amplitude} of soliton
oscillations.
Results based on the 1D GPE, with other parameters as in Fig. 2.
}
\end{figure}

{\it Stabilization Against Decay:} 
For a high energy
soliton, continuous parametric driving
counterbalances damping initially 
(solid line in Fig. 4(a) up to $\omega_{z}t \sim 90$), but
subsequently the soliton starts to decay
($90<\omega_{z}t < 170$), thereby changing  both the amplitude and the phase of the
soliton oscillations \cite{Busch,parker1}. 
The evolution in
the relative phase between
the drive and the soliton oscillations eventually enables the soliton to gain energy
again. After a few such gain-loss
cycles,
the initial deep $v_{0}=0.3c$ soliton finally {\it equilibrates} 
as a shallower soliton of energy $E \approx 35 \hbar \omega_{z}$ 
(corresponding to $v_{0} \approx 0.7c$).

{\it Stabilization at Fixed Energy:}
Some applications may require a soliton to be maintained at a fixed energy. 
For this, one must balance the competing effects of driving and dissipation.
This can be achieved by rephasing the drive relative to the soliton oscillations
at appropriate times, as demonstrated 
by the solid line in Fig. 4(b) for a sequence of four rephasing operations.
In principle, this rephasing can be extended to the duration of the condensate lifetime.
In an experiment, one actually measures the position of the soliton (rather than its energy) \cite{BEC_sol}.
Hence, rephasing could be performed by monitoring the amplitude of the soliton oscillations and adjusting
the drive phase, so that the soliton oscillation amplitude remains constant.
Fig. 4(c) shows the oscillation amplitude of the parametrically driven soliton
(solid black line), which is very similar
to the undamped undriven case (horizontal grey lines), and is clearly distinct
from the undriven dissipative motion (dashed lines), whose amplitude increases until the soliton decays at
$\omega_{z}t = 105$.

Finally, we discuss the relevance of the proposed scheme to current
experiments with atomic BECs. Given a longitudinal confinement 
$\omega_{z}=2 \pi \times 10$ Hz, the presented results correspond to
 $\omega_{\perp}=2 \pi \times 2500$ Hz and a linear `on-axis' density
$n=5 \times 10^{7} (1.5 \times 10^{7})~{\rm m}^{-1}$ of
$^{23}$Na ($^{87}$Rb). The harmonic oscillator time unit is $\tau \approx 15~{\rm ms}$.
In the dissipative example of Fig. 4 with $\gamma=10^{-3}$, this would correspond to a
soliton lifetime of around $1~{\rm s}$, which is consistent with the theoretical predictions for
solitons in highly elongated three-dimensional geometries \cite{Busch,Muryshev2,thermo_inst}.
The paddle beams have a waist $w_{0} = 20~{\rm \mu m}$ and maximum
amplitude $\alpha=7\hbar \omega_{z}$ located around $z_{0}=7l_{z}$.
We have also verified that the results presented here hold for smaller aspect ratios (e.g. $\omega_{\perp}/\omega_{\rm z}=50$). Since, for a given 
aspect ratio,
higher frequencies correspond to faster timescales,
the technique presented here is not sensitive to the
particular soliton lifetime. Increasing, however, $\gamma$ beyond an upper limit  (corresponding to large
dissipation), renders it practically impossible to pump energy into the system.

In summary, we have shown that, in the case of a dark soliton
oscillating in a harmonically trapped Bose-Einstein condensate,
the addition of two out-of-phase
gaussian potentials, with amplitude modulated periodically  
at a fixed frequency close to
the soliton frequency, pumps energy into the
soliton.
This technique, which bears close analogies to parametric driving,
can stabilize the soliton against decay for timescales comparable to the condensate lifetime.
Furthermore, suitable optimization of the phase of the drive can force the soliton to
maintain its initial energy for {\it at least}
a few times its natural lifetime. 
Both effects should be experimentally observable in current
experiments.

We acknowledge discussions with D.J. Frantzeskakis and P.G Kevrekidis and the UK EPSRC for funding.

\end{document}